# Diffusion of Lithium Ions in Lithium-Argyrodite Solid-State Electrolytes from Equilibrium and Nonequilibrium Molecular Dynamics Simulations

Ardeshir Baktash[†], James C. Reid[†], Tanglaw Roman[†,‡,§], Debra J. Searles[*†#]

[†]*Centre for Theoretical and Computational Molecular Science, Australian Institute for Bioengineering and Nanotechnology, the University of Queensland, Queensland 4072, Australia*

[‡]*School of Mathematics and Physics, The University of Queensland, Queensland 4072, Australia*

[§]*School of Physics, The University of Sydney, New South Wales 2006, Australia*

[#]*School of Chemistry and Molecular Biosciences, the University of Queensland, Queensland 4072, Australia*

Email: *d.bernhardt@uq.edu.au*

## Abstract

The use of solid-state electrolytes to provide safer, next-generation rechargeable batteries is becoming more feasible as new materials with greater stability and higher ionic diffusion coefficients are designed. However, accurate determination of diffusion coefficients in solids is problematic and reliable calculations are highly sought-after. In this paper we compare diffusion coefficients calculated using nonequilibrium and equilibrium *ab initio* molecular dynamics simulations for highly diffusive solid-state electrolytes for the first time, to demonstrate the accuracy that can be obtained. Moreover, we show that *ab initio* nonequilibrium molecular dynamics can be used to determine diffusion coefficients when the diffusion is too slow for it to be feasible to obtain them using *ab initio* equilibrium simulations. Thereby, using *ab initio* nonequilibrium molecular dynamics simulations we are able to obtain accurate estimates of the diffusion coefficients of Li ions in $Li_6PS_5Cl$ and $Li_5PS_4Cl_2$, two promising electrolytes for all-solid-state batteries. Furthermore, these calculations show that the diffusion coefficient of lithium ions in $Li_5PS_4Cl_2$ is higher than many other potential all-solid-state electrolytes, making it promising for future technologies. The reasons for variation in conductivities determined using computational and experimental methods are also discussed. It is demonstrated that small degrees of disorder and vacancies can result in orders of magnitude differences in diffusivities of Li ions in $Li_6PS_5Cl$, and these factors are likely to contribute to inconsistencies observed in experimentally reported values. Notably, the introduction of Li-vacancies and disorder can enhance the ionic conductivity of $Li_6PS_5Cl$.

# 1. Introduction

Finding new materials for use in technologies for energy production and storage is one of the challenges that scientists are facing today. In 1991 Sony corporation introduced the first commercial Li-ion batteries as a power source.[1] Since then much research has been carried out to improve the performance and quality of Li-ion batteries.[2-3] Compared to other energy storage materials, Li-ion batteries have shown advantages such as higher energy density and longer life-times.[2, 4] However, after almost three decades of using these batteries, they still have a critical safety problem that is partly due to of the use of organic liquid electrolytes. The flammability and low thermal stability of liquid electrolytes mean batteries with liquid electrolytes can leak or ignite if they become overheated.[5-6] Due to current and future applications of rechargeable batteries in products such as cell phones, laptops, energy storage for short-haul airplanes and electric vehicles, it is very important for industries to find a safe replacement.

In recent years, the all-solid-state battery (ASSB) has been introduced. This type of battery works in the same way as traditional Li-ion batteries, with the most significant difference being in their electrolytes. Instead of a flammable liquid electrolytes, ASSBs have inorganic solid electrolytes with higher thermal stabilities.[7-8] Compared with liquid and polymer electrolytes, ASSBs are safer, lighter, have higher energy density, and are more durable.[9-11] Also, many solid electrolytes are suitable for use in combination with cathode materials with higher potential capacities because of their large window of electrochemical stability.[12]

A solid-state electrolyte that is comparable with its liquid counterpart in terms of performance should be able to conduct lithium ions from the cathode material to the anode material efficiently. In order to achieve this, the ionic conductivity of the electrolyte should be higher than approximately $10^{-3}$ S cm$^{-1}$ at room temperature.[13]

Determination of the conductivity of solid-state electrolytes is complicated in experiments due to the challenges in reproducible synthesis of the materials and the sensitivity of the conductivity on structure. Computational methods provide a resource that can be used to determine conductivity and its dependence on structure and composition of the crystalline materials. In computations, these parameters can be precisely controlled, unlike experimental studies where impurities and defects can be present, which are sensitive to the synthetic conditions and difficult to characterize. However, because of the relatively low conductivity of solid-state electrolyte materials (usually lower than $10^{-3}$ S cm$^{-1}$), in most cases it is very computationally expensive, and sometimes impossible, to directly calculate an accurate value for conductivity of the materials at room temperature using *ab initio* molecular dynamics simulations.[14] One way to solve this problem is to calculate the diffusion coefficient at high temperatures and use the Arrhenius relation to predict a value for ionic diffusion at room temperature.[14-15] However, the relative errors in the final value of the conductivity is likely to be large because statistical errors in the high-temperature data lead to even larger relative errors in the extrapolated results[14] Therefore, new computational methods that can give reliable estimates of the diffusion coefficient need to be applied to study these systems. We show that *ab initio* (AI) nonequilibrium molecular dynamics (NEMD) simulations can be used to this effect. Although AI-NEMD simulations have been used to determine the ionic conductivity of $LiBH_4$, the results obtained were not compared with results from AI equilibrium molecular dynamics (EMD) simulations in the previous work. We show that AI-NEMD simulations enable reliable estimates of the diffusion coefficients to be obtained for materials with diffusivities that are unable to be directly determined using equilibrium calculations. These results can be used to predict materials worthy of consideration as solid-state electrolytes and to identify reasons for variation in experimental measurements.

Among the all-solid-state electrolytes, sulfide-based electrolytes are one of the most promising candidates due to their moderate electrochemical stability, good mechanical properties, and ionic conductivities that are higher than many other potential solid-state electrolytes.[16-19] The Li-argyrodites are a family of sulfide-base electrolytes based on $Li_7PS_6$, some of which have relatively high ionic conductivity of $10^{-5} - 10^{-3}$ S cm$^{-1}$ at room temperature.[20] The Li-argyrodites form a high and low-temperature phase and the ionic conductivity is greater in the high-temperature phase. Most of the higher conductivity argyrodite structures, including $Li_7PS_6$, are not stable at room temperature. However, it has been shown that by making Li vacancies and incorporating halogens into the structure of $Li_7PS_6$, it is possible to form structures with empirical formula $Li_6PS_5X$ (X=Cl, Br and I) that are stable in the higher conductivity phase at room temperature.[21-22] Experiments show that $Li_6PS_5X$ (X=Cl, Br and I) have the same crystallographic structures (space group $F\bar{4}3m$) as $Li_7PS_6$ and the ionic conductivity of $Li_6PS_5Cl$ and $Li_6PS_5Br$ at room temperature is high enough to be considered for battery technology.[23-24] In addition, the effects of halide disorder has been investigated.[22] $Li_6PS_5Cl$ is reported to have a conductivity of ~$10^{-3}$ S cm$^{-1}$ at room temperature and is electrochemically stable up to 7 V vs Li/Li$^+$.[24-25] Calculations have suggested that extra halogens and Li-vacancies result in higher conductivities and it has been proposed that $Li_5PS_4Cl_2$ could be an alternative material, although it has not yet been synthesized.[26]

$Li_6PS_5Cl$ is an argyrodite electrolyte for which there has been much experimental and computational research. However, due to various difficulties mentioned above, the diffusion mechanism in this material is not fully understood and the predictions of the conductivity using computational and experimental results vary over orders of magnitude.[15, 21, 24-27]

The main aims of this paper are to compare AI-NEMD simulations with standard AI-EMD simulations and show that AI-NEMD methods can be used to determine ionic conductivity in low conductivity materials such as all-solid-state electrolytes. All our simulations use *ab initio* molecular dynamics simulations, so we drop 'AI-' from the acronyms AI-EMD and AI-NEMD from this point on. We firstly consider $Li_5PS_4Cl_2$ because it has a lithium ion self-diffusion coefficient that is sufficiently high that EMD simulations can readily be used for its calculation. Thus, the ionic conductivity of $Li_5PS_4Cl_2$ can be determined at different temperatures and using both EMD and NEMD methods. The results are used to demonstrate the accuracy that can be achieved using both methods, and the statistical errors resulting from extrapolation using the Arrhenius equation. The ionic conductivity of $Li_6PS_5Cl$ at room temperature is then determined using NEMD simulations and the results are compared with other computational and experimental results. This material has a much lower diffusion coefficient than $Li_5PS_4Cl_2$, and past results determined using EMD simulations have differed by several orders of magnitude. Finally, the methods are used to study $Li_6PS_5Cl$ with vacancies and defects to help identify the reasons for different experimental reports of the conductivity of $Li_6PS_5Cl$.[15, 21, 24-27]

## 2. Computational Methods

A supercell based on 2 unit cells of $Li_6PS_5Cl$ which has 104 atoms, was considered for all the calculations on this material. The $Li_6PS_5Cl$ argyrodite structure is cubic with space group $F\bar{4}3m$ (space group number 216).[28] In this work the unit cell lattice parameters were determined by energy minimization and were 10.08 Å × 10.08 Å × 10.08 Å. Figure 1 shows the crystal structure of $Li_6PS_5Cl$. In this structure the Li ions occupy 48h Wyckoff positions, and S atoms are distributed in 4a and 4c-sites. Sulfurs in the 4a-sites are bound to the phosphorous atoms (4b-sites) forming $PS_4^{3-}$ (labelled S1), sulfurs as $S^{2-}$ are in 4c-sites and are surrounded by Li ions (labelled S2). The Cl ions are distributed in 4a-sites as well.

The pure and defective structures were formed based on the $Li_6PS_5Cl$ crystal structure. To model the $Li_5PS_4Cl_2$ structure, the S2 sulfur ions were replaced by chloride ions, and to keep the structure charge balanced, one of the Li ions surrounding each S2 was removed from the structure.[26] Details of how the defective structures were modelled are described below.

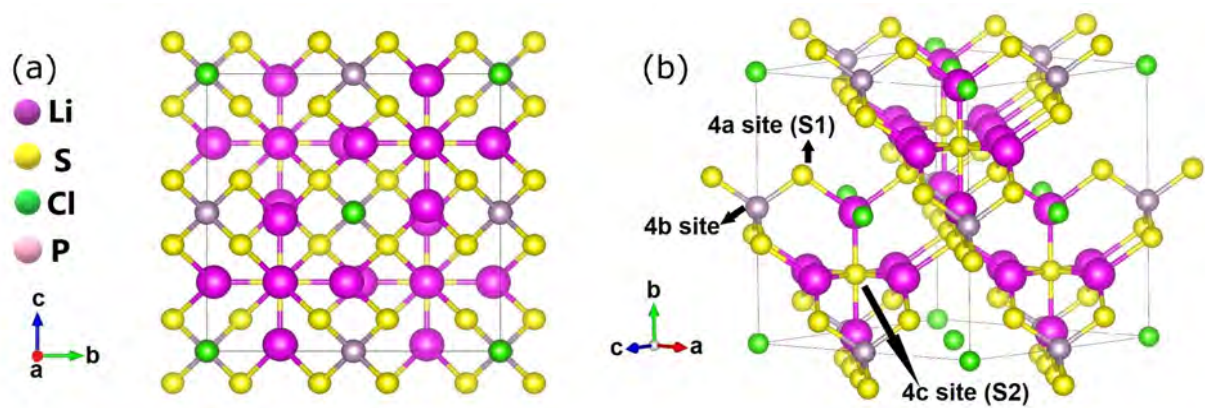

**Figure 1.** Illustration of the crystal structure of $Li_6PS_5Cl$. Li is purple, sulfur is yellow, chlorine is green, and phosphorous atoms, which are in the center of the $PS_4^{3-}$ tetrahedral ion, are light purple.[29]

*Ab initio* Born-Oppenheimer molecular dynamics simulations were performed using the CP2K/Quickstep package and a modified version of this that incorporated the NEMD algorithm discussed below.[30-31] The Perdew-Burke-Ernzerhof (PBE) generalized gradient approximation (GGA)[32] was selected for the DFT exchange-correlation functional. To correct for van der Waals interactions the DFT-D3 [33] method was used. Pseudopotentials of Goedecker, Teter and Hutter (GTH) were employed[34] The DZVP-MOLOPT-SR-GTH [35] basis set was selected because it is optimized for calculating molecular properties in gas and condensed phases. This is a Gaussian and plane-wave (GPW) basis[36] and a cutoff energy of 280 Ry was selected. A 1×1×1 k-point mesh (Γ point) was used in all calculations and preliminary optimization calculations indicated that using grid introduced errors of less than 0.01 Å for lattice parameters and errors of less than 0.06 eV for the energy per unit cell. The optimized lattice constants (10.08 Å) are 2.5% larger than the experimentally reported value for $Li_6PS_5Cl$, and similar to results reported in earlier computational work.[15, 26]

Both EMD and NEMD simulations were used to calculate the diffusion coefficients of the materials in the NVT ensemble. The equations of motion for the EMD simulations are:

$$\dot{\mathbf{q}}_i = \frac{\mathbf{p}_i}{m_i} \qquad (1)$$

$$\dot{\mathbf{p}}_i = \mathbf{F}_i - \alpha \mathbf{p}_i \qquad (2)$$

where $\mathbf{q}_i$ is the position, $\mathbf{p}_i$ is the momentum and $m_i$ is mass of atom $i$, $F_i$ is the interatomic force on atom $i$ and α couples the atoms to the Nosé-Hoover thermostat. A chain thermostat with the chain number of 3 (the default for CP2K) was applied for the EMD calculations:

$$\dot{\alpha} = \frac{1}{Q}\left(\sum_{i=1}^{N}\frac{\mathbf{p}_i^2}{m_i} - gk_BT\right) - \alpha\alpha_2 \qquad (3)$$

$$\dot{\alpha}_2 = \frac{Q\alpha^2 - k_BT}{Q_2} - \alpha_2\alpha_3 \qquad (4)$$

$$\dot{\alpha}_3 = \frac{Q_2\alpha_2^2 - k_BT}{Q_3} \qquad (5)$$

In the above equations $Q$, $Q_1$ and $Q_2$ are the friction coefficients which were all given equal values of 3 in the equilibrium simulations.

The nonequilibrium method used to determine the diffusion coefficients is the color diffusion algorithm which has been widely applied in classical molecular dynamics simulations.[37] Using this algorithm, the equations of motion for the atoms or ions are given by equations (1) and (2), except for the Li ions where the equation of motion for the momentum (2) is modified to include a force due to a color field, $\mathbf{F}_c$:

$$\dot{\mathbf{p}}_i = \mathbf{F}_i + c_i \mathbf{F}_c - \alpha \mathbf{p}_i . \tag{6}$$

Here $c_i$ is the color charge of each of the Li ions. Half of the Li ions have a positive color charge (+1) and the other half have a negative color charge (-1). For NEMD calculations a chain thermostat was not used as it is inappropriate for the nonequilibrium simulations.[38] The equation of motion for the thermostat for nonequilibrium study is therefore given by equation (3) where $\alpha_2$ is equal to zero. For calculations of properties of nonequilibrium systems, the thermostat needs to be applied to the peculiar momentum of the conductive ions. However, because we are extrapolating to zero field, this is not problematic and it is appropriate to use equation (6). The color field forces the Li ions to move in response to the field, but does not change the interatomic interactions (the electric charge on the Li ions is maintained as +1 and contributes to the Coulomb interactions between atoms that is part of $\mathbf{F}_i$). With a color field applied, Li ions with opposite color charges will experience forces that tend to move them in opposite directions through the diffusion channel. The result, on average, will be a color current in the direction of the field. We note that the total momentum of the system is initially zero, and the equations of motion and selection of charges ensure that this is maintained at all times so that the center of mass of the system does not drift.

The EMD and NEMD equations of motions were integrated using a velocity Verlet algorithm with a time step of 1 fs. Where statistical errors are reported, they are one standard error in the mean.

The mean square displacement (MSD) of Li ions was used to calculate the diffusion coefficient from EMD simulations. The self-diffusion coefficient is given by:

$$D_s = \lim_{t \to \infty} \frac{1}{6Nt} \sum_{i=1}^{N} \left\langle \left(\Delta \mathbf{r}_i(t)\right)^2 \right\rangle \tag{7}$$

where $D_s$ is self-diffusion coefficient, $\Delta \mathbf{r}_i(t)$ is displacement of the $i$th of $N$ Li ions over a period, $t$, and $\langle ... \rangle$ indicates an ensemble average. The collective diffusion coefficient $D_c$ (or Li center of mass diffusion coefficient) is given by:

$$D_c = \lim_{t \to \infty} \frac{1}{6Nt} \left\langle \left( \sum_{i=1}^{N} \mathbf{r}_i(t) - \sum_{i=1}^{N} \mathbf{r}_i(0) \right)^2 \right\rangle \tag{8}$$

where $\mathbf{r}_i(t)$ is the position of the $i^{th}$ Li ion at time $t$. The self- and collective diffusion coefficients differ if the ions do not move independently. The conductivity experimentally measured from the application of an electric potential will correspond to the collective diffusion coefficient whereas nuclear magnetic resonance experiments give the self-diffusion coefficient. We note that the statistical error in the computations of the self-diffusion coefficient will be lower than the collective diffusion, so if they are expected to have similar values, it is advantageous to consider the self-diffusion coefficient.

From the self-diffusion coefficient, the conductivity of the material, $\sigma$, can be calculated using the Nernst-Einstein equation:[37]

$$\sigma = \frac{ne^2 Z^2}{k_B T} D_s \tag{9}$$

where $n$ is the ion density of Li, $e$ is the elementary electron charge, $Z$ the valence of Li, $k_B$ is Boltzmann's constant, $T$ is the temperature and $D_s$ is the self-diffusion coefficient of the material at $T$.

Using the NEMD simulations, the self-diffusion coefficient can be determined from the color current produced by the color field. As noted above, this method has previously been used to determine the self-diffusion coefficient of a solid-state electrolyte with low conductivity.[39] The color current is given by:

$$\mathbf{J}_c(t) = \sum_{i=1}^{N} c_i \mathbf{v}_i(t) \tag{10}$$

where $\mathbf{v}_i$ is the velocity of $i$th Li ion. At low fields, the color current in the direction of the field ($J_c = \mathbf{J}_c \cdot \mathbf{F}_c / |\mathbf{F}_c|$) will be linearly proportional to the field, $F_c = |\mathbf{F}_c|$, when the systems is in a steady state, and then:

$$D_s = \frac{k_B T}{N} \lim_{t \to \infty} \lim_{F_c \to 0} \frac{J_c}{F_c} \tag{11}$$

where N is the number of lithium ions subject to a color field. The value of the field below which there is a linear relationship between the color current and applied field will depend on the system and conditions such as temperature. Therefore to use this expression in practice, simulations need to be carried out to determine that critical field. Furthermore, although the color current changes with the color field, in the linear regime the statistical error in the color current does not change. Therefore, to obtain results with the lowest statistical error it is best to use the maximum field for which linear response occurs. For ergodic systems, the ensemble average of the color current in equation (11) can be replaced with a time-average, giving:

$$\langle \mathbf{J}_c \rangle = \lim_{t \to \infty} \frac{1}{t} \int_{t_0}^{t_0+t} \mathbf{J}_c(s)\,ds = \lim_{t \to \infty} \frac{1}{t} \sum_{i=1}^{N} c_i \Delta \mathbf{r}_i(t) \tag{12}$$

So if the field is in the x-direction, the self-diffusion coefficient can also be written as:

$$D_s = \frac{k_B T}{N} \lim_{t \to \infty} \lim_{F_c \to 0} \frac{\sum_{i=1}^{N} c_i \Delta x_i(t)}{t F_c} \quad (13)$$

The field adds a force to the particles in the direction of the field, and has a similar effect to reducing the activation energy barrier for diffusion.[40] If the diffusion process can be modelled as a jump process, the expected time for a single jump will increase exponentially with the size of the activation barrier, and therefore application of the field will allow systems with much lower diffusion coefficients to be considered for a given simulation time. In practice, this means that greater advantage for NEMD calculations is expected for systems where the diffusion coefficient is low.

In this paper, the diffusion mechanism of Li ions in $Li_6PS_5Cl$ and $Li_5PS_4Cl_2$ and the diffusion coefficients/conductivity are studied in detail. Comparison of the EMD and NEMD methods show that both methods can be used to predict the ionic diffusion of these materials when their ionic conductivity is around $10^{-3}$–$10^{-2}$ S cm$^{-1}$. If these materials have much lower ionic conductivity (e.g. at lower temperatures), the EMD simulation times required to get precise and reproducible results are not currently feasible. However, we show that it is feasible to use NEMD to determine diffusion coefficients of solid-state electrolytes with conductivities of around $10^{-6}$–$10^{-4}$ S cm$^{-1}$. In addition, due to the higher ionic conductivity of solid-state electrolytes at higher temperatures, high-temperature ionic conductivities were determined from MD simulations and the Arrhenius relationship between diffusion and temperature was used to estimate the diffusion coefficients of the electrolytes at lower temperatures:

$$D = D_0 e^{-E_a/(k_B T)} \quad (12)$$

For $Li_5PS_4Cl_2$ EMD and NEMD simulations at 300 K, 600 K and 800 K are feasible. In both cases the direct calculation at 300 K is compared with the result from extrapolation of the high temperature-data using the Arrhenius equation. Extrapolation of results from EMD simulations and experiments is widely used to predict the ionic conductivity and barrier energy of solid electrolytes,[15, 41] and this study will allow the suitability of this approach to be assessed for the systems considered.

## 3. Results and discussion

### 3.1. Ionic Conductivity of $Li_5PS_4Cl_2$

The Li ion diffusion coefficient in $Li_5PS_4Cl_2$ was determined using EMD and NEMD simulations and the results were compared. The EMD simulations at 300 K, 600 K and 800 K were run for 45, 40 ps and 20 ps, respectively. The results of these EMD simulations were used to obtain the average MSD of the Li ions. The results for 10 replicas at each temperature, and the statistical errors at each temperature were obtained. Figures 2 (a), (b) and (c) show the MSD of the Li ions as a function of time at 300, 600 and 800 K for 10 independent trajectories. For diffusive motion, the MSD increases linearly with time at long times, and the slope is related to the diffusion coefficient through equation (7). figure 2 (d), (e), and (f) show the average MSD of the ten replicas including the error bars from which the self-diffusion coefficient and the corresponding conductivity of the material can be determined with an estimate of the precision of the calculation. The results are shown in Table 1.

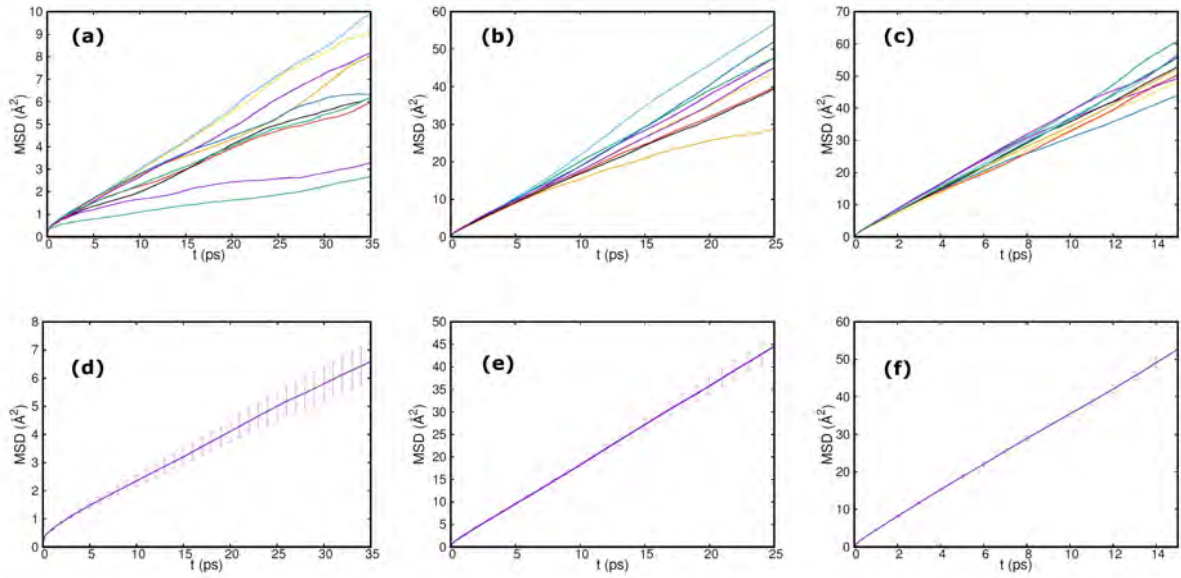

**Figure 2.** Mean square displacement (MSD) of Li ions as a function of time for 10 replicate simulations of Li$_5$PS$_4$Cl$_2$ at (a) 300 K, (b) 600 K and (c) 800 K, and average MSD of the ions as a function of time for 10 replicates with the error bars at (d) 300 K, (e) 600K and (f) 800 K.

**Table 1.** The self-diffusion coefficient and corresponding conductivity of Li$_5$PS$_4$Cl$_2$ at different temperatures calculated using EMD and NEMD simulations. The numbers in brackets refers to errors in the last decimal place.

| Temperature | EMD | | NEMD | |
|---|---|---|---|---|
| (K) | $D_s$ (cm$^2$ s$^{-1}$) | $\sigma$ (S cm$^{-1}$) | $D_s$ (cm$^2$ s$^{-1}$) | $\sigma$ (S cm$^{-1}$) |
| 300 | 2.9 (4) × 10$^{-6}$ | 0.35 (5) | 3.3 (4) × 10$^{-6}$ | 0.40 (5) |
| 600 | 2.9 (2) × 10$^{-5}$ | 1.8 (1) | 2.9 (2) × 10$^{-5}$ | 1.8 (1) |
| 800 | 5.6 (3) × 10$^{-5}$ | 2.5 (1) | 5.2 (4) × 10$^{-6}$ | 2.4 (2) |

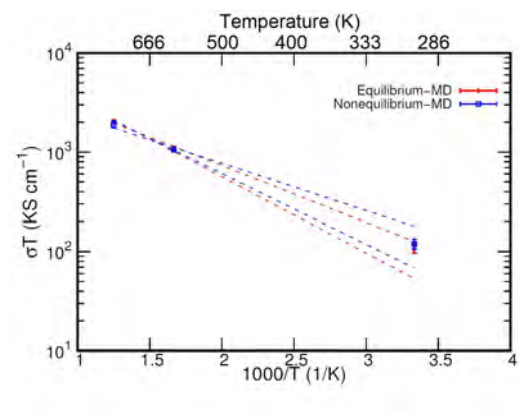

**Figure 3**. Arrhenius plot of the conductivity for $Li_5PS_4Cl_2$ from EMD simulations (red) and NEMD simulations (blue) at various T (data points). Red lines (EMD) and blue lines (NEMD) show the bounds of error for estimation of extrapolated data at 600 K and 800 K to 300 K assuming Arrhenius behavior.

For solid-state electrolytes that have low conductivity at room temperature (~$10^{-3}$ S cm$^{-1}$ or lower) it is computationally expensive to directly calculate a precise value for the conductivity at room temperature. To solve this issue it is common to calculate the conductivity of the material at higher temperatures and extrapolate the data to room temperature to give an estimate of the conductivity.[15, 41] An Arrhenius plot of the values of the conductivity calculated from the EMD simulations of the Li ion in $Li_5PS_4Cl_2$ from the MSDs at different temperatures is shown in figure 3 (red data points). Furthermore, to demonstrate how extrapolated results compare with the values that are directly determined, the predicted results obtained from extrapolation of the higher temperature data (600 and 800K) are shown. Based on the MSD calculations, $Li_5PS_4Cl_2$ has an ionic conductivity of $0.35 \pm 0.05$ S cm$^{-1}$ at 300 K. figure 3 shows that by considering error bars at 600 and 800 K, the calculated conductivity using extrapolation has a large statistical error, giving a result of between 0.17 and 0.37 S cm$^{-1}$ at 300 K. This is in agreement with the directly calculated value. Although the statistical errors for the higher temperature results are small compared to the values, the relative errors at lower temperatures are large due to the extrapolation, as can be seen from figure 3.

To determine the ionic conductivity of solid-state electrolytes Aeberhard et al.[39] used an alternative approach that has been widely applied in classical simulations, but had not been used in *ab initio* simulations, and is based on NEMD simulations as described above. They determined the and calculated the self-diffusion coefficient of hexagonal $LiBH_4$ at 535 K. In the present work, to show the accuracy of NEMD simulations the method was used to predict the self-diffusion coefficients and conductivities of $Li_5PS_4Cl_2$ at 300 K, 600 K and 800 K, and the results are compared with the results determined from the EMD MSD calculations.

In order to calculate the ionic conductivities from NEMD simulations, it is necessary to determine the conductivity at several fields to identify the linear response regime. At each field in the linear regime, 10 simulations were carried out for 15 ps, 13 ps and 5 ps at 300 K, 600 K and 800 K, respectively. These times were selected to give a total computational effort similar to that required for the EMD simulations. Figures 4 (a) and (b) show the time integral of the color current due to a color field of $F_c = 0.04$ eV Å$^{-1}$ for $Li_5PS_4Cl_2$ at 600 K using the methodology described above. To identify the linear regime, at least two points with similar values for the conductivity or the diffusion coefficient are required. Figure 4 (c) shows the time integral of the color current due to color fields of 0.02 and 0.03 eV Å$^{-1}$ at 300 K. Figure 5 shows the time-averaged color current density of the Li ions as a function of field strength at 300, 600 and 800 K.

The Arrhenius plot for the conductivities calculated using NEMD simulations at different temperatures is presented in Table 1 and shown by the blue data in figure 3. The values obtained by the EMD and NEMD methods agree to within the limits of error (within one standard error of the mean). The conductivity of the material at 300 K is also estimated by extrapolating from higher temperatures. Comparing the red and blue lines in figure 3, it is clear that using similar total simulation times for NEMD and EMD calculations at 300, 600 and 800 K, the statistical errors in the NEMD simulation results were almost the same as the statistical errors of the results from the EMD method. This was true for the results obtained for direct measurement at 300 K and from extrapolation of the high temperature results to 300 K. These outcomes are consistent with the fact that for this highly diffusive system, the lithium ions are readily able to move between different regions of the sample in all cases and a field is not required to force the ions from one 'cage' to the next.

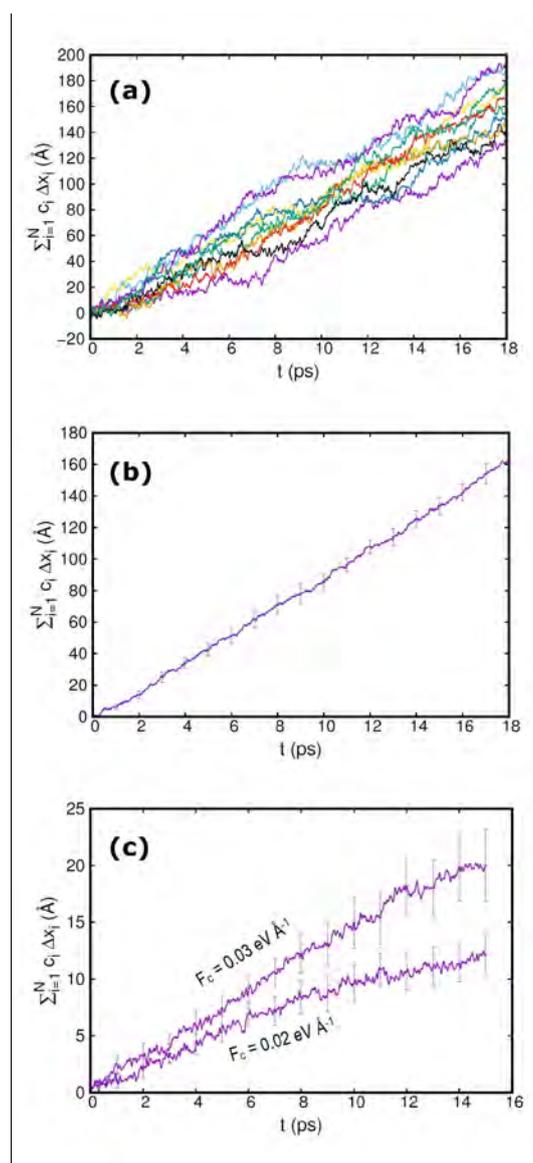

**Figure 4.** (a) Time-integral of the Li ion color current in $Li_5PS_4Cl_2$ due to a color field as a function of time for 10 independent simulations at 600 K using NEMD simulations and a color field $F_c = 0.04$ eV . (b) The average over 10 independent runs of the time-integral of the color current of the Li ions at 600 K and color field $F_c = 0.04$ eV $Å^{-1}$. (c) Time-integral of the color current due to a color field as a function of time for a field of $F_c = 0.02$ eV $Å^{-1}$ and $F_c = 0.03$ eV $Å^{-1}$ for Li ions in $Li_5PS_4Cl_2$ at 300 K.

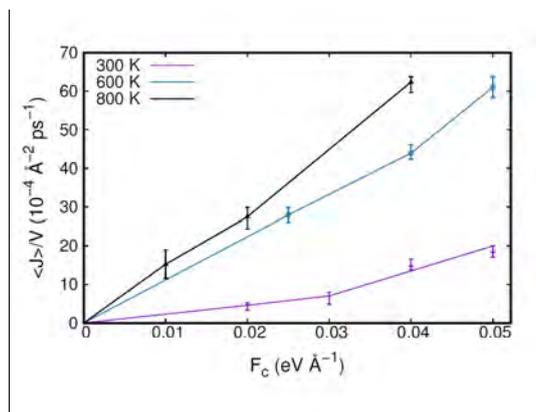

**Figure 5.** The time-averaged color current density vs the external color field strength for Li ions in $Li_5PS_4Cl_2$ at 300, 600 and 800 K. The lines provide a guide to the eyes.

### 3.2. Collective diffusion coefficient of $Li_5PS_4Cl_2$

The Nernst-Einstein equation is often used to relate the conductivity to the self-diffusion coefficient. However, if the diffusing atoms or molecules in the studied sample are not moving independently during diffusion then this might not be an adequate approximation and the collective diffusion coefficient should be considered.[42] The self and collective diffusion coefficients will be the same if the diffusing atoms or molecules move independently of each other, but they will differ otherwise (for example, if they move as a cluster). We therefore used equation (8) to calculate the collective diffusion coefficient for $Li_5PS_4Cl_2$ and compare this to the self-diffusion coefficient to check that this approximation is adequate for the systems we consider. The most diffusive pure system we considered in this manuscript was used for this purpose. It was selected because the statistical error is much larger for the collective diffusion calculations than for the self-diffusion calculations and it would be difficult to draw conclusions if the statistical error is too large.

Figure 6 compares the MSD of the ions and their center of mass for $Li_5PS_4Cl_2$ at 800K. Note that the ion MSD was determined from 10 independent runs and the ion center of mass MSD from 30 independent runs (each run for 20 ps), yet the error bars for the ion MSD remain much smaller. This is because each of the ions in the sample could provide an independent contribution to the MSD. It is clear that the MSD calculated in both ways agree to within the limits of error for this material and therefore the self and collective diffusion coefficients will agree. The agreement is consistent with independent behavior of single Li ions in $Li_5PS_4Cl_2$ during the diffusion process and therefore gives us confidence that for this system the conductivity can be calculated using the self-diffusion coefficient.

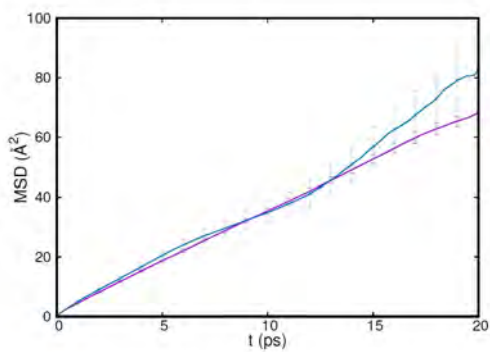

**Figure 6.** MSD (purple) and center of mass MSD (blue) as a function of time for Li ions in $Li_5PS_4Cl_2$ at 800 K

### 3.3. Ionic Conductivity of $Li_6PS_5Cl$

Since simulation times for diffusive motion scale approximately inversely with the diffusion coefficient,[43] in systems with low conductivity it is expected that EMD simulations would need to be very long to obtain accurate results, and at some stage no movement of ions between regions or cages in the electrolyte will be observed in a feasible time-scale. From the calculations on $Li_5PS_4Cl_2$, it was found that combining NEMD simulations at high temperatures with extrapolation to low temperatures could give accurate results. Therefore, we propose that this could be a way of extending the range of materials for which the conductivity can be calculated by providing a method when the conductivity is so low that EMD simulations cannot be used. Here we test this proposal by calculating the ionic conductivity of $Li_6PS_5Cl$ using NEMD simulations.

Both EMD and NEMD simulations were carried out with the aim of calculating the conductivity and understanding the mechanism of diffusion in pure $Li_6PS_5Cl$, $Li_6PS_5Cl$ with S-Cl disorder (S and Cl swapping positions) and $Li_6PS_5Cl$ with both $Li^+$ vacancies and S-Cl disorder.

### 3.3.1. Pure $Li_6PS_5Cl$

We can distinguish between two types of motion in this system. One where a set of 6 Li ions move in an octahedral region about an S2 sulfur atom and another where the Li ion moves (jumps) between these octahedral regions (or cages). If the timescale for the jumps is longer than the simulation time, then a diffusion coefficient cannot be determined from the simulation. Using EMD simulations of $Li_6PS_5Cl$, no jumps between cages were observed in 10 independent trajectories of 100 ps at 300 and 450 K and very few jumps were observed for some of the trajectories at 600 K. This suggests a very low diffusion coefficient of the material at 300 K (less than ~$10^{-4}$ S cm$^{-1}$) and demonstrates the difficulty of measuring the value of the conductivity through EMD simulations.[43] However, it was possible to obtain the conductivity values using NEMD simulations at both 600 K and 800 K, where jumps could be observed, and the conductivity at 300 K was estimated using extrapolation of these results, assuming Arrhenius behavior. In figure 7, the red unfilled squares show the calculated conductivities at 600 and 800K (the error bars are smaller than the symbols), and the lines show the bounds of predictions for the conductivity of pure $Li_6PS_5Cl$ at lower temperatures. As can be seen from figure 7, the conductivity of the pure $Li_6PS_5Cl$ is predicted to be $10^{-5}$ – $10^{-4}$ S cm$^{-1}$. The simulation time required to obtain similar precision for the conductivity of $Li_6PS_5Cl$ using the EMD method at 600 and 800 K was prohibitive, and we were not even able to estimate a mean and standard error because jumps did not occur. An alternative would be to consider higher temperatures but the extrapolation errors would be greater than if this was necessary, and it could lead to a disruption of the structure and non-Arrhenius behavior for the diffusion coefficient. Therefore the NEMD approach is more appropriate in this case.

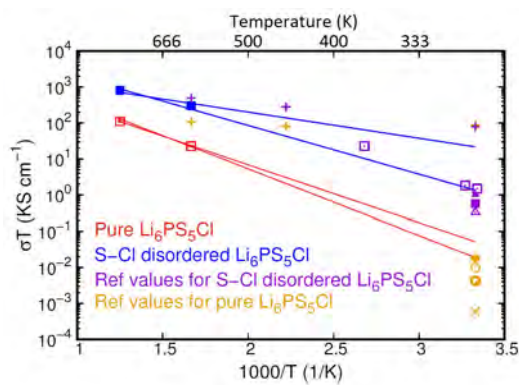

**Figure 7.** Data for the Li ion conductivity in pure (red squares and lines) and S-Cl disordered (blue squares) $Li_6PS_5Cl$ from this work. The lines show bounds for the predicted Li ion conductivity in pure (red lines) and S-Cl disordered (blue lines) $Li_6PS_5Cl$ obtained by extrapolation, assuming Arrhenius behavior for the diffusivity. The other points are computational and experimental results from the literature. The cross (x)[15] and the plus (+) sign[26] are computational results and filled[44] and unfilled[44] squares, filled[45] and unfilled[24] triangles, filled[46] and unfilled[28] circles and the bold unfilled circle [27] are experimental results. For statistical errors in the results (where available), see Table 2.

The Li ion conductivity for pure $Li_6PS_5Cl$ has previously been determined experimentally[27-28, 46] and computationally[15, 26] at 300 K and these results are also shown in figure 7 as circles (experimental data) and crosses (computational data). It can be seen there are differences in the orders of magnitude of results at 300 K. Of particular note is that the computational results from the literature for pure $Li_6PS_5Cl$ at 300 K differ by 5 orders of magnitude and are between one and four orders of magnitude different from the experimental results. The lower value was obtained by extrapolation of data for simulations at 600 K and higher. Therefore it seems that there are no reliable computational estimates of the diffusion coefficient for this system prior to the current work. Our computational prediction is similar to the highest experimental result. The statistical error in the computational results from the previous studies were not reported in most cases, but is expected to be high based on the computational time and supercell sizes considered, and the inherent error propagation when extrapolating the data for materials with low conductivity. It is clear that for this system, the low jump frequency of Li ions in $Li_6PS_5Cl$ at room temperature (~$10^9$ s$^{-1}$ according to ref.[28]) requires extrapolation of the higher temperature to predict the conductivity at lower temperatures and use of NEMD.

The experimental results differ by a factor of two, and it was proposed in the literature that this could be due to the annealing temperature of the samples which causes different degrees of Cl and S disorder,[47] the level of the crystallinity,[24, 48] existence of extra chlorine in the synthesized structure, Li vacancy in the structure,[48] and existence of impurities[44] of the synthesized samples. The effects of some of these factors are considered in the sections below.

### 3.3.2. S-Cl disorder in $Li_6PS_5Cl$

It has been proposed that the high conductivity of $Li_6PS_5Cl$ observed in some experiments could be due to disorder of Cl and S atoms.[28, 44, 48] To test and understand this, we consider a model of the structure with disorder, based on the experimentally observed structure. In the structure, 25% of the Cl ions that are occupying 4a sites in $Li_6PS_5Cl$ are exchanged with S2 ions which are occupying 4c sites. Data from EMD simulations were then used to determine the conductivity of the disordered structure at 600 K and 800 K. The results are shown in figure 7 as filled blue squares (the error bars are smaller than the symbols). Assuming Arrhenius behavior, the results are extrapolated to lower temperatures as shown by the blue lines in figure 7. Experimental results for the conductivity of the disordered structure are shown in figure 7 as squares and triangles[24, 44-45, 47] and a reported computational value[26] is shown as a plus sign (+). Considering the error bars, our calculated value for the conductivity of disordered structure agrees with the recent experimental reports of Yu et al.[44] According to our results, the conductivity of the disordered structure is approximately two orders of magnitude higher than the pure structure. Therefore it is clear that this apparently small change in the structure which involves swapping 4 of the 104 atoms, results in a very significant change in the conductivity. This indicates that slight defects in the pure crystal could lead to very different experimental conductivities.

The diffusion pathways of the Li ions in the ordered and disordered structures at 600 K were also determined using EMD simulations and are shown in figure 8. Comparing the pathways, it is apparent that introducing disorder by swapping the positions of Cl and S ions significantly changes the motion of the Li ions. As seen in figure 8 (a) and (b), in the pure systems Li ions move inside the octahedral cages formed by the $PS_4^{3-}$ (see figure 1) and the absence of pathways between the cages indicate that the energy barrier for diffusion out of the cages is high compared to the thermal energy available, and consequently the ionic conductivity of the pure crystal is too low to be determined from these EMD simulations. However, in the disordered structure a channel between the cages is evident in the right half of the structure (where the ions are swapped) in figures 8 (c) and (d). This indicates that in the direction of the new channel, the barrier energy for Li ion diffusion to the other cages is much smaller than it was in the pure structure and Li ions move in the disordered part of the structure relatively easily. The presence of a channel with a lower energy barrier explains the increase in the ionic conductivity of the disordered $Li_6PS_5Cl$. Also, variation in the concentration of S2-Cl disorder and dispersion of the disordered sites could explain the different experimentally reported conductivities.

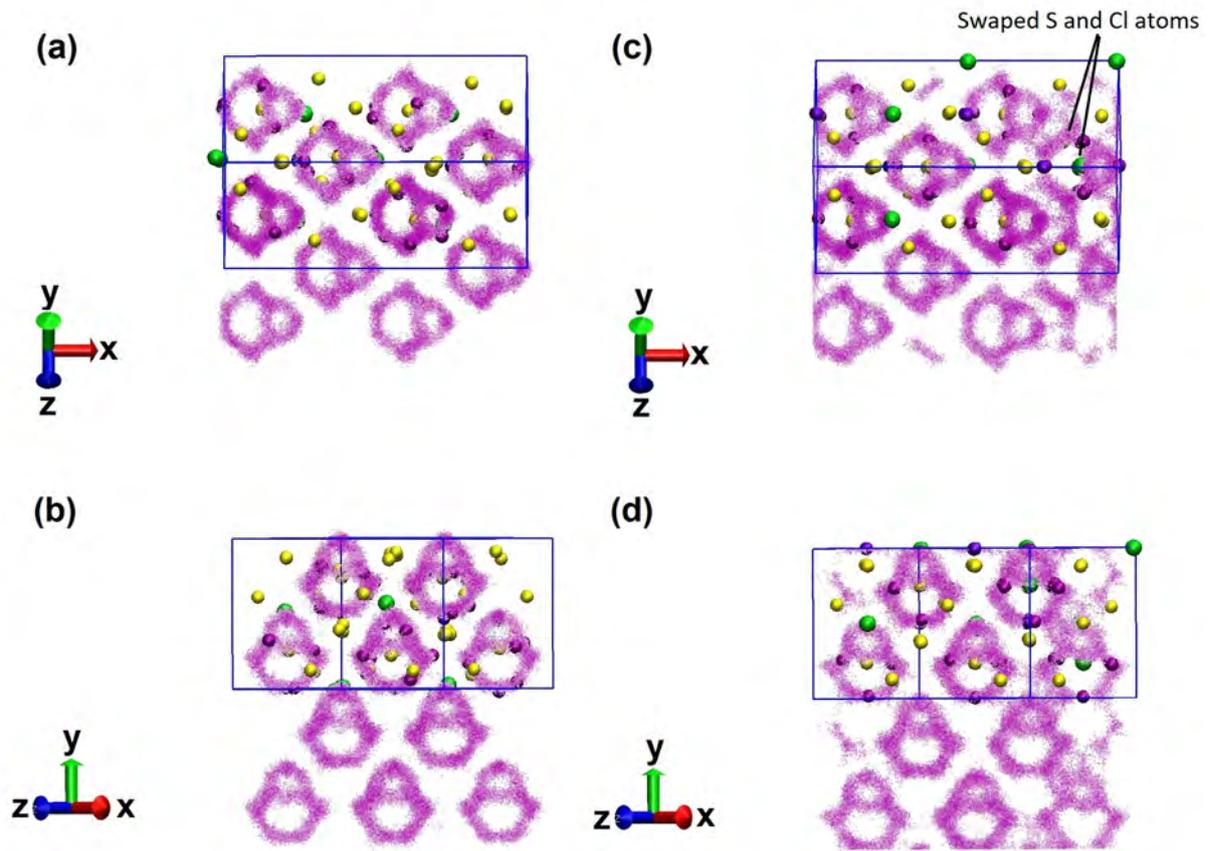

**Figure 8.** Different views of Li trajectories in (a), (b) pure and (c),(d) S-Cl disordered $Li_6PS_5Cl$ at 600 K monitored for 50 ps. The violet, yellow, light purple and green atoms are lithium, sulfur, chlorine and phosphorus, respectively. The small violet dots show the diffusion pathways of the lithium ions. The boxes show one simulation supercell.

### 3.3.3. Effect of Li vacancies in S-Cl disordered $Li_6PS_5Cl$

It has been proposed that Li vacancies can be present in samples of $Li_6PS_5Cl$ which would be expected to increase their Li ion conductivity.[48] In a study considering this,[48] the experimentally synthesized sample was suggested to have experimental formula $Li_{5.6}PS_{4.8}Cl_{1.2}$ after annealing. We note that the charges are not balanced in the proposed empirical formula, $Li_{5.6}PS_{4.8}Cl_{1.2}$. However, a similar formula with charge balance is $Li_{5.8}PS_{4.8}Cl_{1.2}$. To study the effect of the combination of a Li-vacancy, extra Cl ions and fewer S2 ions[47] on the ion conductivity of the $Li_6PS_5Cl$, we remove two Li ions from the $Li_6PS_5Cl$ supercell (8 unit cells) and replace two S2 ions with Cl ions[49] giving a new structure of $Li_{5.75}PS_{4.75}Cl_{1.25}$, which is similar to the empirical formula given in the literature ($Li_{5.8}PS_{4.8}Cl_{1.2}$). The ionic conductivity of this structure at 300 K was calculated to be $0.09 \pm 0.03$ S cm$^{-1}$, which is several magnitudes higher than the pristine $Li_6PS_5Cl$ solid electrolyte ($6 \times 10^{-5}$ - $3 \times 10^{-4}$ S cm$^{-1}$) and higher than the estimated value for $Li_6PS_5Cl$ with disorder of the S2 and Cl atoms only ($6 \times 10^{-3}$ - $1 \times 10^{-1}$ S cm$^{-1}$). The estimated value for the ionic conductivity is also higher than that reported in ref.[48] ($1.1 \times 10^{-3}$ S cm$^{-1}$), but the sensitivity to vacancies and disorder means that these would not be expected to agree due to the difference in empirical formulas. Clearly, Li vacancies and substitution of an S ion with a Cl ion significantly changes the diffusion in the structure. Table 2 presents a summary of conductivities of the different type $Li_6PS_5Cl$ structures (pure, disordered and structures having both disorder and Li vacancies) from experimental and computational reports.

Based on the results from our calculations and previously reports,[15, 21, 24, 26-28, 48] it can be concluded that the conductivity of pure $Li_6PS_5Cl$ is relatively low ($10^{-5}$ – $10^{-4}$ S cm$^{-1}$). However, impurities like Li-vacancies, grain boundaries, and ion disorder introduced during their synthesis would affect the final conductivity of the synthesized argyrodite electrolyte.

**Table 2.** The conductivity of $Li_6PS_5Cl$ and defective materials at 300 K determined in various experiments and computational studies including this work.

| Material | Conductivity / S cm$^{-1}$ | Source |
| --- | --- | --- |
| $Li_6PS_5Cl$ | $1.4 \times 10^{-5}$ | Experiment [27] |
| | $3.3 \times 10^{-5}$ | Experiment [28] |
| | $6 \times 10^{-5} – 3 \times 10^{-4}$ (a) | This work (NEMD) |

| | 6×10⁻⁵ | Experiment [46] |
|---|---|---|
| | 0.29 | Computation (MSD) [26] |
| | 0.05 (0.16) | Computation (Jump) [26] |
| | 2×10⁻⁶ | Computation (MSD) [15] |
| $Li_{5.6}PS_{4.8}Cl_{1.2}$ | 1.1×10⁻³ | Experiment [48] |
| $Li_{5.75}PS_{4.75}Cl_{1.25}$ | 6 ×10⁻² – 1.2 ×10⁻¹ [a] | This work |
| $Li_6PS_5Cl$ with Cl and S2 disorder | 1.9×10⁻³ | Experiment [47] |
| | 4.96×10⁻³ | Experiment [44] |
| | 3.38×10⁻³ | Experiment [45] |
| | 0.26 | Computation (MSD) [26] |
| | 0.89 (1.29) | Computation (Jump) [26] |
| | 6 ×10⁻³ – 1×10⁻¹ [a] | This work (EMD) |

[a] The ranges given are based on extrapolation of data at 800 and 600 K, for which error bars of one standard error are assumed.

## 4. Conclusions

In this paper we used EMD and NEMD simulations to determine the ionic conductivity of solid electrolytes and the effect of disorder and defects on diffusion of the Li ions. Agreement of the results from these methods showed that they are both able to provide reliable estimates of conductivity when the conductivity is sufficiently high. However, for solid-state electrolyte with low conductivity, which are common at temperatures closer to room temperature, it is necessary to use NEMD simulations. The advantage of NEMD over EMD calculations increases as the diffusivity decreases. This is because the time required to explore a material in equilibrium simulations will inversely proportional to the diffusion coefficient [43] whereas the applied field will also contribute to this in NEMD simulations.

We note, that if the diffusion is not isotropic it will be necessary to carry out several NEMD simulations with fields in different directions in order to obtain the diffusion coefficients. Therefore the efficiency of the NEMD calculations would be reduced in this case. However, it will still be advantageous, or necessary, at sufficiently low fields.

We also studied the conductivity of $Li_5PS_4Cl_2$ and $Li_6PS_5Cl$ in detail. Based on the results from our calculations, $Li_5PS_4Cl_2$ is predicted to be a highly conductive solid electrolyte although it has not yet been synthesized. Based on our results for $Li_6PS_5Cl$, although the pure material has a relatively low ion conductivity ($6 \times 10^{-5} - 3 \times 10^{-4}$ S cm$^{-1}$ at 300 K) we confirmed that by increasing Li vacancies of the structure or introducing disorder in the ionic positions of the Cl and S ions, it is possible to enhance the ion conductivity of this structure. Although these systems had been studied previously, the size of the error bars made it difficult to ascertain the effects. It can be concluded that the higher experimentally reported conductivity of $Li_6PS_5Cl$ could be due to combinations of Li ion vacancies and Cl-S ion disorder or maybe higher concentration of halogen (Cl) after annealing. We note that our computational results either predict conductivities that are higher than the experimental results, or on the high end of the range of experimental results. There are some systematic errors in the computations that might contribute to this including the system size that can be modelled, the level of theory used in the *ab initio* molecular dynamics simulations and changes in the lattice parameters during diffusion which is carried out at under constant volume conditions. However, the presence of grain boundaries, impurities, disorder and the inhomogeneous distribution of disordered sites in the experimental samples could also explain the differences between the experimental and computational results. Nevertheless, the computational results are reproducible and the trends due to changes in the structure indicate ways in which a material can be tuned to increase the conductivity which is crucial for the improvement of solid-state-electrolytes.

# Acknowledgments

The authors thank the Australian Research Council for support of this project through the LIEF and Discovery programs (LE0882357, LE160100051 and DP140100193). The authors also thank Dr Qinghong Yuan for her comments and helpful discussions. We acknowledge access to computational resources at the NCI National Facility through the National Computational Merit Allocation Scheme supported by the Australian Government. This work was also supported by resources provided by the Pawsey Supercomputing Centre with funding from the Australian Government and the Government of Western Australia. We also acknowledge support from the Queensland Cyber Infrastructure Foundation (QCIF) and the University of Queensland Research Computing Centre.